\documentclass[10pt,conference]{IEEEtran}

\usepackage{research5IEEE}
\usepackage{psfrag}
\usepackage{graphicx}

\usepackage{amsmath, amssymb}

\newtheorem{lemma}{Lemma}
\newtheorem{corollary}{Corollary}
\newtheorem{theorem}{Theorem}
\newtheorem{remark}{Remark}

\newtheorem{definition}{Definition}

\newcommand{\bfpi}{\boldsymbol{\pi}}
\begin{document}

\title{On Cognitive Interference Networks}

\author{
\authorblockN{Amos Lapidoth}
\authorblockA{ETH Zurich \\
CH-8092 Zurich, Switzerland \\
Email: lapidoth@isi.ee.ethz.ch}
\and
\authorblockN{Shlomo Shamai (Shitz)}
\authorblockA{Technion,
Haifa\\
Haifa 32000, Israel  \\
Email: sshlomo@ee.technion.ac.il}
\and
\authorblockN{Mich\`ele A. Wigger }
\authorblockA{ETH Zurich \\
CH-8092 Zurich, Switzerland \\
Email: wigger@isi.ee.ethz.ch}
}

\maketitle 

\begin{abstract}
  We study the high-power asymptotic behavior of the sum-rate capacity
  of multi-user interference networks with an equal number of
  transmitters and receivers. We assume that each transmitter is
  cognizant of the message it wishes to convey to its corresponding
  receiver and also of the messages that a subset of the other
  transmitters wish to send. The receivers are assumed not to be able
  to cooperate in any way so that they must base their decision on the signal they
  receive only. We focus on the network's pre-log, which is defined as the
  limiting ratio of the sum-rate capacity to half the logarithm of the
  transmitted power.

  We present both upper and lower bounds on the network's pre-log. The
  lower bounds are based on a linear partial-cancellation scheme which
  entails linearly transforming Gaussian codebooks so as to eliminate
  the interference in a subset of the receivers. 
  
  \emph{Inter alias}, the bounds give a complete characterization of
  the networks and side-information settings that result in a full pre-log,
  i.e., in a pre-log that is equal to the number of transmitters (and
  receivers) as well as a complete characterization of networks whose
  pre-log is equal to the full pre-log minus one. They also fully
  characterize networks where the full pre-log can only be achieved if
  each transmitter knows the messages of all users, i.e., when the
  side-information is ``full''.% The bounds also characterize the
 % networks where side-information increases the pre-log only when the
 % side-information is full.
\end{abstract}

\section{Introduction}\label{sec:Intro}

In this paper we study communication scenarios that arise in wireless
networks when multiple spatially-separated transmitters communicate
to multiple spatially-separated receivers. %Our scenario can be viewed as a generalization of the
%two-user interference channel to more then two transmitters and
%receivers. We shall therefore refer to the considered scenario as an
%interference network.

Consider a situation where $K$ non-cooperating
transmitters, labeled $\{1,\ldots, K\}$, wish to communicate with $K$ non-cooperating
receivers, labeled $\{1,\ldots, K\}$, where  Receiver~$j$
wants to learn Message $M_j$ for each $1\leq j\leq K$. 
 Here $\{M_j\}_{j=1}^K$ are independent with $M_{j}$ being uniformly
distributed over the set $\{1,\ldots, \lfloor e^{n R_j}\rfloor\}$, 
where $n$ denotes the block-length of transmission and
$R_j$ is the rate of transmission to Receiver~$j$.
 
We assume that each transmitter is cognizant of  a subset of  the messages
$\{M_1,\ldots, M_K\}$ and denote the set of indices of the messages
 known to Transmitter $k$  by $\set{S}_k$, $k \in
\set{K}=\{1,\ldots, K\}$.  
Also, we assume that the labeling of the
transmitters is such that Transmitter~$k$ knows Message~$M_k$ and hence 
$\{k\}\subseteq \set{S}_k\subseteq \set{K}$. Transmitter~$k$
computes its sequence of inputs at times $1$ to $n$,
$\vect{X}_k^n\triangleq\trans{(X_{k}(1),\ldots, X_{k}(n))}$ as a function
of the set of Messages $\{M_j\}_{j\in \set{S}_k}$.

A  setting where every transmitter
knows all the involved messages---i.e., where $\set{S}_k=\set{K}$ for
all $k\in\set{K}$---will be called the \textit{full side-information}
setting, and a setting where every Transmitter~$k$ is
cognizant only of the Message~$M_k$---i.e., where
$\set{S}_k=\{k\}$ for all $k \in\set{K}$---will be called the \textit{no
  side-information} setting. The \textit{full
  side-information} setting is also called ``fully cognitive network'',
and it corresponds to a broadcast channel with multiple
receivers. The \textit{no side-information} setting is also
called ``non-cognitive network'' and is a generalization of the
two-user interference channel to more than two transmitters
and more than two receivers. A network with neither \textit{full
  side-information} nor \textit{no side-information} is called a
\textit{partial side-information} network.  We will refer to any of
the above settings as interference  networks.

The interference networks are described by a fixed channel matrix
$\mat{H}\in \Reals^{K\times K}$, where $\Reals$ denotes the set of
real numbers, as follows. Denote the  output
signals observed at Receivers 1 through $K$  at the discrete time-$t$
by $Y_{1}(t)$ through
$Y_{K}(t)$. The output vector at time-$t$
$\vect{Y}(t)\triangleq\trans{(Y_{1}(t), \ldots, Y_{K}(t))}$ is given
by
\begin{equation}\label{eq:channel}
\vect{Y}(t)  =  \mat{H} \vect{x}(t)+ \vect{Z}(t), \quad 1\leq t \leq n,
\end{equation}
where $\vect{x}(t) \triangleq\trans{(x_{1}(t),\ldots, x_K(t))}$ is the
time-$t$ input vector consisting of the inputs at Transmitters 1
through $K$, and where $\{\vect{Z}(t)\}$ is a sequence of independent
and identically distributed (IID) Gaussian random vectors of zero-mean
and covariance matrix $\mat{I}_K$. (Here $\mat{I}_K$ denotes the
identity matrix of dimension $K$.) Throughout the paper the channel
matrix $\mat{H}$ is assumed to be of full rank.

For each transmitter we impose the same average block power constraint on the
sequence of channel inputs, i.e.,
\begin{equation}\label{eq:power}
\frac{1}{n} \E{ \sum_{t=1}^{n} X_{k}^2(t)} \leq P, \quad k\in \set{K}.
\end{equation}

We say that a rate-tuple $(R_1,\ldots, R_K)$ is achievable if there
exists a sequence of pairs of encoding schemes satisfying
\eqref{eq:power} and decoding schemes such that in the limit as $n$
tends to infinity the probability of a decoding error at each receiver
tends to 0. Note that each receiver bases its decision on the signal
it receives only. Denoting by $R_\Sigma$ the sum of the rates
$R_1,\ldots, R_K$, i.e.,
\[ R_\Sigma =\sum_{j=1}^K R_j\] 
 we can define the sum-rate capacity
$C_{\Sigma}(P, \mat{H}, \{\set{S}_k\})$ as the
supremum of the sum-rates over all achievable rate tuples.

In this work we focus on the behavior of the sum-rate capacity
$C_\Sigma(P ,\mat{H} ,\{\set{S}_k\})$ in the high SNR regime, i.e., in
the limit when $P \rightarrow \infty$. In particular, the quantity of
interest in this regime is the limit of the ratio of the sum-rate
capacity to the Gaussian single-user channel capacity when the
available power tends to infinity:
\begin{equation}\label{eq:def_pre-log}
\eta\left(\mat{H}, \{\set{S}_k\} \right)\triangleq \varlimsup_{P\rightarrow \infty} \frac{C_{\Sigma}(P, \mat{H}, \{\set{S}_k\})}{\frac{1}{2}\log\left( 1+
  P\right)}.
\end{equation}
The limiting ratio $\eta\left(\mat{H}, \{\set{S}_k\} \right)$
determines the logarithmic growth of the sum-rate capacity at high
power, and we will refer to it as the pre-log of the network.  Note
that the pre-log depends both on the message sets $\{\set{S}_k\}_{k\in
  \set{K}}$ and on the channel matrix $\mat{H}$. The main goal of this
work is to examine the influence of the sets $\{\set{S}_k\}_{k\in
  \set{K}}$ on the pre-log of an interference network with given
channel matrix $\mat H$.
  
For \textit{full side-information} settings the pre-log is already
known to be equal $K$ \cite{yoogoldsmith06}.  However, for
\textit{partial side-information} settings and for \textit{no
  side-information} settings the pre-log is not yet known for general
interference networks. But see \cite{etkintsewang07},
\cite{jafarfakhereddin07}, \cite{devroyesharif07}, and
\cite{lapidothshamaiwigger07} for some special networks.
   
In \cite{etkintsewang07} the
two-transmitters/two-receivers interference network with \textit{no
  side-information} is investigated. The results therein include
the result that the  pre-log of the
setting  equals 1 and furthermore even characterize the capacity region
of the network to within 1 bit.

The pre-log of the two-transmitters/two-receivers network with \textit{partial side-information}
was studied in \cite{devroyesharif07}. There it was shown that for no \textit{partial
  side-information} setting the pre-log is larger than 1; only
\textit{full side-information} yields the ``full'' pre-log 2.

The more general scenario
where both transmitters and both receivers can communicate with multiple
antennas is treated in \cite{jafarfakhereddin07}.

It should be emphasized that our setting does not include as a special
case the X-channel where each transmitter sends independent messages to
the \emph{two} receivers \cite{maddahetal06,devroyesharif07}.

%In contrast to this it has been shown in \cite{maddahetal06} and
%\cite{devroyesharif07} that in a 2-transmitters/2-receivers
%X-channel---i.e., in a scenario where each transmitter sends
%independent messages to both receivers%, and each transmitter is not
%cognizant of the messages of the other transmitter
%% ---the pre-log in
%% the \textit{no side-information} setting can be larger than the
%% pre-log in the corresponding interference network, where each
%% transmitter sends only a message to one of the receivers. Also, there
%% are \textit{partial side-information} settings on the X-channel with a
%% pre-log which is larger than the pre-log of the \textit{no
%%   side-information} setting. Thus, the specific distribution of the
%% messages at the transmitters characteristic for X-channels can
%% increase the pre-log compared to interference network settings.

In contrast to the described works in this submission we consider
networks with generally more than two transmitters and receivers.

Recently, the authors \cite{lapidothshamaiwigger07} considered a
particular example of an interference network with more than two
transmitters and more than two receivers. They  showed that 
in interference networks \textit{partial side-information} settings
can exist with a larger pre-log than in the \textit{no
  side-information} setting. In particular, the authors considered an
interference network where the channel matrix is given by the matrix
with ones on the diagonal, some constant $\alpha$ on the first lower
secondary diagonal, and 0 everywhere else. Thus, in the considered
network, Receiver~$j$ observes the sum of Transmitter~$j$'s input
signal, Transmitter~$(j-1)$'s input signal scaled by the factor
$\alpha$, and additive white Gaussian noise. For this network it was
shown that partial side-information can increase the pre-log
significantly and even lead to the ``full'' pre-log $K$, the same
pre-log as in the \textit{full side-information} setting.

Thus, we see that for the two-transmitters/two-receivers interference
  network described in \cite{devroyesharif07} and for the interference
  network described in \cite{lapidothshamaiwigger07} the impact of \textit{partial
  side-information} on the pre-log is drastically
  different.  This fact might not seem so surprising to the reader
  since the two networks have a very different structure. However, it
  is not clear which properties of a network determine how
  \textit{partial side-information} influences the pre-log. In
  fact we will show later in this paper that for the two
  similar networks with channel matrices 
\begin{equation}\label{eq:H1}
\mat{H}_1=\begin{pmatrix}1 & 1/2 & 1/4\\ 1/2 &1 &1/2\\
0&1/2 &1  \end{pmatrix}
\end{equation}
and 
\begin{equation}\label{eq:H2}
\mat{H}_2=\begin{pmatrix}1 & 1/2 & 0\\ 1/2 &1 &1/2\\
0&1/2 &1  \end{pmatrix}
\end{equation}
the dependence of the pre-log  on the message sets $\{\set{S}_k\}$ is  completely
different. For networks with channel matrix $\mat{H}_1$ in the \textit{no side-information} setting  the pre-log equals 1, and there are
\textit{partial side-information} settings with pre-log equal 2
and \textit{partial side-information} settings with pre-log
3. In contrast, for networks with channel matrix $\mat{H}_2$, in any
\textit{partial side-information} setting and in the \textit{no
  side-information} setting the pre-log equals 2 and only in the \textit{full
  side-information} setting the pre-log equals 3.

%In the next section we will identify which properties of
%a network determine how \textit{partial side-information}
%  influences the pre-log of  a setting.
%of the pre-logs depending on the amount of available
%side-information

\section{Main Contributions}\label{sec:main}
In this section we state the main results of our work. For  proofs
we refer to a forthcoming longer version of this paper.

We begin by
stating for which interference network settings we can determine the
pre-log exactly based on the lower bound and
the upper bound derived in the last two subsections. In the second subsection we  characterize when \textit{partial
  side-information} can increase the pre-log, and in the third subsection we
give some examples of specific networks to illustrate the results in
the previous two subsections.
In the subsection before last we describe an encoding
scheme---the \emph{linear partial-cancelation} scheme--- leading to the lower
bound on the pre-log. Finally, in the
last subsection we describe how to derive the upper bound on the pre-log.

\subsection{Exact Results}
For general
interference settings there is a gap between the upper bound
 and 
the lower bound obtained with a linear partial-cancelation scheme. Nevertheless, for certain networks the two bounds meet,
thus demonstrating the asymptotic optimality of the linear
partial-cancelation scheme. 
Examples of
such settings include the setting described in
\cite{lapidothshamaiwigger07} and also---for any given message sets $\{\set{S}_k\}$---the
fully connected 2-by-2 interference networks and the networks with
channel matrix $\mat{H}_2$ given in \eqref{eq:H2}. For \textit{no
  side-information} settings and for certain \textit{partial
  side-information} settings the bounds also meet for networks with channel
matrix $\mat{H}_1$ given in \eqref{eq:H1}. Also, the lower bound and the upper bound
also meet for all settings where $p^*=K-1$ and (trivially) where
$p^*=K$. Here $p^*$, which is given ahead in \eqref{eq:p1}, is the
best pre-log achieved with a linear partial-cancelation scheme.

From the lower bound and the upper bound we obtain the
following results on the pre-log $\eta(\mat{H},\set{S}_k)$ depending on
$p^*$.
\begin{theorem}\label{th:main}
Consider an interference network with channel matrix $\mat{H}$ and message
sets $\{\set{S}_k\}$. Let $p^*$ be defined as in \eqref{eq:p1}. Then:
\begin{eqnarray}
p^*=K &\Longrightarrow& \eta(\mat{H},\{\set{S}_k\}) =K, \label{eq:p1}\\
p^* =  K-1 &\Longrightarrow &\eta(\mat{H},\{\set{S}_k\}) = K-1,\label{eq:p2}\\
p^*  \leq  K-2 &\Longrightarrow &\eta(\mat{H},\{\set{S}_k\}) < K-1
.\label{eq:p3}
\end{eqnarray}
\end{theorem}
 Since $p^*$ takes on only positive integer values smaller or equal to
 $K$, the following corollary can be obtained from
 Theorem~\ref{th:main}.
\begin{corollary}\label{th:corr}
For an interference network with channel matrix $\mat{H}$ and message
sets $\{\set{S}_k\}$:
\begin{equation}
\eta(\mat{H},\{\set{S}_k\})=K \Longleftrightarrow p^*=K,
\end{equation}
and 
\begin{equation}
\eta(\mat{H},\{\set{S}_k\})=K -1\Longleftrightarrow p^*=K-1.
\end{equation}
Furthermore, the pre-log $\eta(\mat{H},\set{S}_k)$ can never take
value in the open interval $(K-1,K)$.
\end{corollary}
This result is somewhat surprising since for certain interference networks the
pre-log can indeed be a non-integer value. An example of an
interference network with non-integer pre-log is given in Section
\ref{sec:enc}.

\subsection{When Partial Side-Information increases the Pre-log}
With the results of Theorem~\ref{th:main} in mind we address the
following two problems: the problem of  identifying
 the channel matrices $\mat{H}$ for which \textit{full
   side-information} is necessary in order
 to have ``full'' pre-log $K$; and the problem of identifying the channel matrices $\mat{H}$ for
 which  \textit{partial
 side-information} is  beneficial, in the sense that there is a
 \textit{partial side-information} setting with a pre-log which is
 larger than the pre-log of the
 \textit{no side-information} setting. The following theorem answers
 these questions.
\begin{theorem}\label{th:partial-SI} Consider an interference network
  with channel matrix $\mat{H}$ and let $\mat{H}_{(j)}^{(k)}\in
  \Reals^{(K-1) \times (K-1)}$ denote  the matrix
obtained  when deleting the $j$-th row and the
$k$-th column from the channel matrix $\mat{H}$, and let 
$h_{j,k}$ denote the element of $\mat{H}$ in row $j$ and column
$k$. Then
\begin{enumerate}
\item The message sets $\{\set{S}\}_{k\in\set{K}}$ have to fulfill the
  following sufficient and necessary conditions for the
  pre-log to equal $K$:
\begin{IEEEeqnarray*}{rCl}
&\left(  \eta(\mat{H},\{\set{S}_k\})=K\right)&\\& \Longleftrightarrow& \\
 &\left( \forall j,k \in \set{K}:
\quad \left(\textnormal{rank}\left(
      \mat{H}_{(j)}^{(k)}\right)=K-1 \Longrightarrow j\in
    \set{S}_k\right) \right).& 
% j\in \set{S}_k\text{ whenever
% }\textnormal{rank}\left(\mat{H}_{(j)}^{(k)}\right)=K-1, \quad j,k\in\set{K} \hspace{1cm}
% \nonumber \\
%\Longleftrightarrow \hspace{1cm}\quad \quad  \eta(\mat{H},\{\set{S}_k\})=K .\hspace{1cm}
\end{IEEEeqnarray*}
Thus, in particular, \textit{full side-information} is necessary  for that the
  pre-log
  of a network $\mat{H}$ is equal $K$, if and only if,
  $\textnormal{rank}\left(\mat{H}_{(j)}^{(k)}\right)=K-1$ for all
  $j\neq k$, and $j,k\in
\set{ K}$.
\item Let $\set{H}$ be the union of the set of all diagonal $K$-by-$K$
  matrices and  of the set of all $K$-by-$K$  matrices for which there is an index
  $k^*\in \set{K}$ such that
\begin{equation}\label{eq:condhkj}
h_{j,k}=\begin{cases} 0, & \text{if } j\neq k , \quad j\neq k^*, k \neq k^*
\\
\text{arbitrary, } & j=k=k^*  \\ \neq 0 , & \text{else }% j=k^* \text{ or }
 % k=k^*  \text{ or } j=k \neq k^* , \quad j, k \in \set{K}\\
\end{cases}.
\end{equation}
%\begin{IEEEeqnarray}{rCl}\label{eq:condhkj}
% h_{k,j} =h_{j,k}=0,\quad  j\neq k \neq k^*, j, k \in \set{K}
% \nonumber \\
% h_{k,k^*}\neq 0, \quad   h_{k^*,k}\neq 0,\quad h_{k,k}\neq 0,
%\quad k\neq k^*, k \in \set{K}.\nonumber \\%\IEEEeqnarraynumspace %\nonumber \\
%\end{IEEEeqnarray}
Then, for all channel matrices $\mat{H}$ in $ \set{H}$ 
the pre-log of any \textit{partial side-information} setting  equals the
pre-log of the \textit{no side-information} setting. \\
For all channel matrices which are not contained in  the set $\set{H}$ there exist \textit{partial side-information} settings with a pre-log
which is strictly
  larger than the  pre-log of  the \textit{no
    side-information} setting.
\end{enumerate}
\end{theorem}
In the remaining of this section we want to have a closer look at
Conditions~\eqref{eq:condhkj}. The matrices satisfying these
conditions can be illustrated as follows:
{\small \begin{equation}\label{eq:H}
\mat{H}= \begin{pmatrix} 
\times& 0 & 0 & \ldots & 0 & \times &  0 & \ldots & 0 & 0 \\ 
0   & \times &0 & \ldots & 0& \times &   0 & \ldots & 0 &0 \\ \\
 &&&\ldots &&& \ldots\\\\
0   & 0&0 & \ldots & \times& \times&   0 & \ldots & 0&0 \\
\times & \times & \times & \ldots &\times & ?&\times &  \ldots &
\times& \times \\
0   & 0&0 & \ldots & 0& \times &   \times & \ldots & 0&0 \\\\
 &&&\ldots &&& \ldots\\\\
 0 & 0& 0& \ldots & 0 &\times & 0 & \ldots &\times & 0\\
 0 & 0& 0& \ldots & 0 &\times & 0 & \ldots &0 & \times    \end{pmatrix}
\end{equation}}
where the index of the row with $K-1$ occurrences of ``$\times$'' is the same as the
index of the column with $K-1$ occurrences of ``$\times$''. At all
positions which are marked by  an ``$\times$'' the matrix $\mat{H}$
must contain a non-zero element, but these elements do not have to be
identical. At the position which is marked by ``?'' the matrix
$\mat{H}$ can be arbitrary, possibly also 0. 

\begin{remark}\label{rem:1}The pre-log of
interference networks with channel matrices
of the form given in \eqref{eq:H}
 equals $K-1$ in the \textit{no side-information} setting and in any
\textit{partial side-information} setting, and the pre-log   equals
$K$ only in the
\textit{full side-information} setting.
\end{remark}
\subsection{Examples}\label{sec:Examples}
\subsubsection{ The fully
connected 2-by-2 interference network}
The two-transmitters/two-receivers  interference network  with channel matrix with only
non-zero components is of the structure illustrated in
\eqref{eq:H}. Thus  with Remark~\ref{rem:1} it is possible
to reconstruct the results about the interference network in
\cite{devroyesharif07}, that is,  that the  pre-log equals 2 only  in the \textit{full
  side-information} setting whereas in the \textit{partial
  side-information} setting the  pre-log equals 1, the same as in the \textit{no
  side-information} setting. 

\begin{remark}
The  fully connected
2-by-2 interference network and trivially the single-user channel are
the only fully connected interference networks---i.e., networks with a channel matrix with only non-zero
components---for which there is no \textit{partial side-information} setting with
pre-log larger than the pre-log of the  \textit{no side-information} setting.
\end{remark}

\subsubsection{Networks $\mat{H}_1$ and $\mat{H}_2$}
Next, let us  consider again the channel matrices $\mat{H}_1$ and
$\mat{H}_2$. We see that  the channel matrix $\mat{H}_2$ is of the
form displayed in \eqref{eq:H} and therefore, by Remark~\ref{rem:1} we can conclude---as
announced in Section~\ref{sec:Intro}---that in any \textit{partial
  side-information} setting and in the \textit{no side-information}
setting the pre-log equals 2 and in the \textit{full side-information}
setting the pre-log equals 3.

For the channel matrix $\mat{H}_1$ we see that the sub-matrix 
\[\mat{H}_{1,{(3)}}^{(1)}=\begin{pmatrix} 1/2 & 1/4 \\ 1 &
  1/2\end{pmatrix}\]
is of rank $1$. Therefore, we can conclude that the interference
network with channel matrix $\mat{H}_1$ and message sets
$\set{S}_1=\{1,2\}$ and 
$\set{S}_2=\set{S}_3=\set{K}$  has pre-log 3.
%\[\eta( \mat{H}_1, \{ \set{S}_1=\{1,2\}, \set{S}_2=\set{K},
%\set{S}_3=\set{K}\})=3.\]
Since all other sub-matrices of the form $\mat{H}_{1,(j)}^{(k)}$ for
$j\neq k$ and $(j,k)\neq (3,1)$ have rank $K-1$, we can also conclude
that in any other \textit{partial side-information} setting the
pre-log is at most 2. Furthermore, computing the rates achievable with
the linear partial-cancelation scheme %using the algebraic characterization
%of $p^*=2$ given in \eqref{eq:pstar} 
one easily finds the message sets
$\{\set{S}_k\}$ such that %of the
%\textit{partial side-information} settings for which 
$p^*=2$ and hence
$\eta(\mat{H}_1, \{\set{S}_k\})=2$.  In the \textit{no
  side-information} setting  with channel matrix $\mat{H}_1$
the pre-log is given by $\eta\left(\mat{H}_1,
  \{\set{S}_k=\{k\}\}\right)=1$. This follows from the upper
bound in
Lemma~\ref{th:convpartial}.% in Section~\ref{sec:main}.
% From Theorem~\ref{th:partial-SI} one also directly obtains that  for 
%  the 3-by-3 interference network with full-rank channel matrix
%  \[
%\mat{H}=\begin{pmatrix} 1 & 1/2 &0 \\ 1/2 & 1
%  &1/2  \\ 0& 1/2&  1  \end{pmatrix} \]
%% for  any kind of \textit{partial
%  side-information} setting and for the \textit{no side-information}
%  setting the pre-log equals 2 whereas for the \textit{full
%  side-information} setting the pre-log equals 3.

\subsubsection{Wyner's Linear Cellular Interference Model}

In \cite{wyner94} Wyner introduced a linear model for cellular
wireless communication systems. The network model is a
symmetric version of the network considered 
in \cite{lapidothshamaiwigger07}, this is, a $K$-by-$K$
interference network where Receiver~$j$
observes the sum of Transmitter~$j$'s input signal,
Transmitter~$(j+1)$'s input signal scaled by a factor $\alpha \neq 0$,
Transmitter~$(j-1)$'s input signal scaled by the same factor $\alpha$, and
additive white Gaussian noise. Thus the channel matrix is given by 1's
on the main diagonal, $\alpha$'s on the first upper and lower
secondary diagonals and 0 every where else.

In his work Wyner considered the case when all receivers are
allowed to cooperate, and hence the setting becomes a multi-access
setting. Here, we consider the case where the
receivers are not allowed to cooperate, and we also assume that the
transmitters have some kind of side-information about the other
transmitter's messages. More precisely, let each transmitter beside its own
message know the messages of the $J$ previous transmitters and the
messages of the $J$ next transmitters for some integer $J \geq
0$. 

%With the results from Section~\ref{sec:main} we obtain the pre-log of
The pre-log of this setting for given parameters $\alpha, J$, and $K$
can be shown to be
\begin{equation*}
\eta_{\textnormal{Wyner}}(\alpha, J, K)= K -\left \lfloor
  \frac{K}{J+2} \right \rfloor.
\end{equation*}
Note that the functional dependence of the pre-log for this setting on the parameters
$\alpha, J$, and $K$ is the same as the functional dependence of the
pre-log for  the asymmetric setting in \cite{lapidothshamaiwigger07}
on these parameters.

\subsection{A Lower Bound}\label{sec:enc}
We propose an  encoding scheme---the linear
partial-cancelation scheme---for an
arbitrary
interference network with channel matrix $\mat{H}$ and message sets
$\{\set{S}_k\}$ as described in Section~\ref{sec:Intro}. The encoding
scheme is based on random coding arguments. 

Prior to transmission, $K$
independent random codebooks $\set{C}_1,\ldots, \set{C}_K$ are
generated according to a zero-mean Gaussian
distribution of variance $P$. Here,  the codebook $\set{C}_j$ is the
set of $n$-length codewords $\left\{\vect{u}_j^n(1), \ldots, \vect{u}_j^n\left(\lfloor
e^{n R_j}\rfloor\right)\right\}$, and it is used to encode the Message $M_j$,  $j\in \set{K}$. Then,
the codebooks are revealed to all transmitters and to all receivers. 

For the encoding each
 transmitter forms a linear combination of the codewords
 $\vect{u}_j^n(M_j)$ where it
 knows Message $M_j$ and such that the input power constraint
 \eqref{eq:power} is satisfied. Thus, Transmitter $k$'s input sequence
% $\vect{X}_k=\trans{(X_k(1), \ldots, X_k(n))}$  
is given by
\begin{equation*}
\vect{X}_k^n = \sum_{j\in\set{S}_k} d_{j,k} \vect{u}_j^n(M_j), \quad k\in \set{K},
\end{equation*} 
for some real coefficients $d_{j,k}$ satisfying 
\[
\sum_{j\in\set{S}_k} d_{j,k}^2 \leq 1, \quad k \in \set{K}.
\]
For every choice of coefficients
$\{d_{j,k}\}_{k \in \set{K}, j\in\set{S}_k}$ we can define the set $\set{R} (\{d_{j,k}\})\subseteq
\set{K}$ of all indices $j$ such that the interference for
Receiver~$j$ is canceled.  
More precisely, the set $\set{R} (\{d_{j,k}\})$  is the set of all
$j\in\set{K}$ such that the
received sequence $\vect{Y}_j^n= \trans{( Y_{j}(1),\ldots, Y_{j}(n))}$  at
Receiver~$j$ can be expressed as 
%More precisely, the set $\set{R} (\{d_{j,k}\})$ is the set of maximum cardinality fulfilling that 
%for some  non-zero real numbers $\{\xi_j\}_{j\in \set{R}(\{d_{j,k}\})} $  the
%received sequence $\vect{Y}_j= ( Y_{j}(1),\ldots, Y_{j}(n))$ at Receiver~$j$ is given by
\begin{equation}\label{eq:interference_cancel}
\vect{Y}_j^n = \xi_j \vect{u}_{j}^n(M_j) + \vect{Z}_{j}^n, \quad j \in \set{R}(\{d_{j,k}\})
\end{equation}  
for $\xi_j\neq 0$.
Note that the set $\set{R} (\{d_{j,k}\})$ depends on the channel matrix $\mat{H}$, on
the message sets $\{\set{S}_k\}_{k\in\set{K}}$,  and of course also on
the chosen coefficients $\{d_{j,k}\}$.

Let 
\begin{equation}\label{eq:p1}
p^*(\mat{H},\{\set{S}_k\})=\max_{\{d_{j,k}\}} |\set{R}(\{d_{j,k}\})|,
\end{equation}
where $|\set{A}|$ denotes the cardinality of the set $\set{A}$.
%For a given network and given message sets we are interested in finding
%a set $\set{R^*}$ of maximum cardinality for which there exists a valid
%choice of coefficients $\{d_{j,k}\}$ 
%such that $\set{R}
%(\{d_{j,k}\})=\set{R}^*$, so that  the interference  is canceled at all receivers with index in
%$\set{R^*}$. We denote this maximum cardinality  by $p^*$. 
%Notice that by choosing $\{d_{j,k}\}$ such that $\set{R}
%(\{d_{j,k}\})=\set{R}^*$ 
If $\{d_{j,k}^*\}$ achieves $p^*(\mat{H},\{\set{S}_k\})$, then by using $\{d_{j,k}^*\}$ the
original interference network is
transformed into $p^*(\mat{H},\{\set{S}_k\})$ parallel Gaussian single-user channels and a
network with $K$ transmitters and $K-p^*(\mat{H},\{\set{S}_k\})$ receivers.
Since on the parallel Gaussian single-user channels the rates $\frac{1}{2}\log
\left( 1+  \xi_j^2 P\right)$ are achievable,  $ j\in \set{R}(\{d_{j,k}^*\})$,
the following lower bound on the sum-rate capacity is obtained 
\begin{equation}\label{eq:csumlow}
C_{\Sigma}(P, \mat{H}, \{\set{S}_k\}) \geq \frac{p^*(\mat{H},\{\set{S}_k\})}{2} \log \left(1+
  \min_{j\in\set{R}(\{d_{j,k}^*\})}\xi_j^2 P\right),
\end{equation}
and hence 
\begin{equation}\label{eq:preloglow}
 \eta(\mat{H},\{\set{S}_k\})\geq p^*(\mat{H},\{\set{S}_k\}).
\end{equation}
%In the sequel we will refer to the described encoding scheme as
%\emph{linear partial-cancellation scheme}.

Inspired by \cite{weingartenshamaikramer07}, we can improve
the linear partial-cancelation scheme by extending it over $\mu>1$ consecutive
channel uses.
To this end, 
let the encoder and the decoder group $\mu$ consecutive channel
uses into a single channel
use of a new $K$-by-$K$ multi-antenna interference network
where each transmitter and each receiver consists of $\mu$ antennas.
Note that any achievable rate tuple for the new network is also achievable, when
divided by $\mu$, on the original network.  As we next show, we can
derive an achievable tuple for the new network by introducing linear
processing at the receivers; by converting it to a new $\mu K$-by-$\mu
K$ single-antenna interference network; and by then applying the
linear partial-cancelation scheme to the
resulting network.
%% We want
%% to transform the obtained multi-antenna $K$-by-$K$ interference network into a single-antenna $\mu K$-by-$\mu K$
%% interference network as described in Section~\ref{sec:Intro} so that
%% we can apply the linear partial-cancelation scheme to the $\mu
%% K$-by-$\mu K$ network.

We split Message $M_j$, $j\in\{1,\ldots,K\}$,
into $\mu$ independent Sub-Messages $M_{(j,1)}\ldots,
M_{(j,\mu)}$ such that there is a one-to-one mapping between $M_{j}$
and the tuple $(M_{(j,1)},\ldots, M_{(j,\mu)})$.\footnote{For example one can think of this splitting
as describing the original message $M_j$ by a sequence of bits and
then splitting up this sequence into disjoint (not necessarily equally long)
bit-sequences and let every sub-message be described by a different
sub-sequence.} 

%With this definition we are ready to identify a $\mu K$-by-$\mu K$
%single-antenna interference network to which we can apply the linear
%partial-cancelation scheme. But---similarly to the scheme described in
%\cite{weingartenshamaikramer07}---first we want to introduce an
%additional degree of freedom to the resulting coding scheme. 
As in \cite{weingartenshamaikramer07} we let
Receiver~$j$ of the multi-antenna $K$-by-$K$ interference network
linearly process the observed $\mu$ antenna outputs by multiplying
them with an arbitrarily chosen $\mu$-by-$\mu$ matrix $\mat{A}_{j}$.
%(It is this linear processing that makes for the possibility
%that when applying the linear partial-cancelation scheme over this
%extended network a higher pre-log can be achieved  than with the
%simple linear partial-cancelation scheme over the original channel.)
The network is now converted to a single-antenna $\mu K$-by-$\mu K$ network
treating the $\mu K$ outputs of  the linear processings
$\mat{A}_1,\ldots, \mat{A}_K$ as %receive antennas as 
separate receivers and by
treating  each $\mu$-tuple of transmit antennas as corresponding to
$\mu$ single users that are cognizant of each others messages.

Indexing the transmitters and receivers of the $\mu K$-by-$\mu K$
network by $(k,i)$ and $(j,i)$ respectively where $1\leq k, j\leq K$
and $1\leq j\leq \mu$ we can describe the network as follows: The
message sets are 
\begin{equation*}
\set{S}_{(k,i)} =\left \{ (k',i'):k'\in \set{S}_k, 1\leq i'\leq \mu \right \} 
\end{equation*}
and the channel matrix
\begin{equation*}
\mat{H}_\mu(\mat{H},\{\mat{A}_j\}) = \left( \mat{H}\otimes
  \mat{I}_\nu\right) \textnormal{diag}\left(\mat{A}_1,\ldots, \mat{A}_K\right)
\end{equation*}
where $\otimes$ denotes the Kronecker product and where
diag$(\mat{A}_1,\ldots,\mat{A}_K)$ denotes the block-diagonal matrix
with the blocks $\mat{A}_1,\ldots,\mat{A}_K$.
If the
rate-tuple $(R_{(j,1)},\ldots, R_{(j,\mu)})$ is achievable in the
$\mu K$-by-$\mu K$ interference network then the rate 
\begin{equation}\label{eq:ratesn}
R_j=\frac{1}{\mu}
\left(R_{(j,1)}+
\ldots +R_{(j,\mu)}\right)
\end{equation}
 is achievable in the original interference network. %% The scaling by
%%  $1/\mu$ arises since one channel use of the $\mu K$-by-$\mu K$
%%  interference network corresponds to $\mu$ channel uses of the
%%  original channel.
 
For the  described $\mu K$-by-$\mu K$ interference network we can apply the linear partial-cancelation
scheme and hence we obtain the achievability of the pre-log:
$p^*(\mat{H}_{\mu}\left(\mat{H}, \{\mat{A}_j\}\right),\{\set{S}_{(k,i)}\})$.
Combined with \eqref{eq:ratesn} this yields a bound on the pre-log
of the original network:
\begin{equation*}
\eta( \mat{H}, \{\set{S}_k\}) \geq \frac{p^*(\mat{H}_{\mu}\left(\mat{H},\{\mat{A}_j\}\right),\{\set{S}_{(k,i)}\})}{\mu}
\end{equation*}
for any set of processing matrices $\{\mat{A}_j\}$.
Hence the best lower bound on the pre-log one can obtain by extending
the linear
partial-cancelation over several channel uses is given by
\begin{equation}\label{eq:preloglow_best}
\eta( \mat{H}, \{\set{S}_k\}) \geq \sup_{\mu \in \Integers^+}
\max_{\{\mat{A}_j\}_{j\in\set{K}}} \frac{p^*(\mat{H}_{\mu}\left(\mat{H},\{\mat{A}_j\}\right),\{\set{S}_{(k,i)}\})}{\mu}.
\end{equation}

That this
modification of the linear-partial cancelation scheme indeed leads
to an improvement in the achievable rates (and in the lower bound
on the pre-log) over the rates (and over the lower bound on the pre-log)
achieved in the original linear partial-cancelation
scheme can be seen in the following example.%%  Also, we show that for any $\mu >1$ there exists a network
%% such that we  need to extend the linear
%% partial-cancelation scheme over $\mu$ channel uses
%%  before increasing the pre-log compared to the simple linear
%% partial-cancelation scheme

\subsubsection{Extending the Linear Partial-Cancelation Scheme over
  several Channel Uses helps}
In this section we want to give an example of a network where by
extending the linear partial-cancelation scheme over several channel
uses leads to a pre-log which is strictly larger than the pre-log
achieved with the simple linear partial-cancelation scheme. 

Consider the family of channel matrices $\{\mat{H}_K\}$ indexed by
the number of transmitters and receivers $K$. For a given $K>1$ %the matrix $\mat{H}_K$ is
%a real $K$-by-$K$ matrix and consists of all 1's except for the first lower
%secondary diagonal and the top right corner element
%the matrix equals 0. Thus, 
we consider a $K$-by-$K$ interference network
where Receiver~$j$, $j\in\{1,\ldots,K\}$, receives a noisy version of the sum of all input
signals except for that of Transmitter~$(j-1)$ where $j-1$ should be
interpreted as $K$ when $j=1$.
%% the input signal of the $(j-1)$-th transmitter, corrupted
%% by additive white Gaussian noise. Receiver~$1$ receives the sum of all
%% input signals, except the input signal of Receiver~$K$, corrupted by
%% additive white Gaussian noise. We consider the \textit{no
%%   side-information} setting, this is, $\set{S}_k=\{M_k\}$ for
%% $k \in \{1,\ldots,K\}$.

%% With the result presented in Section~\ref{sec:main} we obtain that
The
pre-log of the described settings is given by
\begin{equation*}
\eta(\mat{H}_K , \{\set{S}_k=\{M_k\}\}) = \frac{K}{K-1}, \quad
 K>1.
\end{equation*}
To show that  this pre-log is indeed achievable the linear
partial-cancelation scheme needs to be extended over $K-1$ channel
uses. Extending the scheme to less than $K-1$ channel uses achieves
only a pre-log of 1.

\subsection{An Upper Bound}\label{sec:conv_res}
In this section   we provide an upper bound on the sum of the rates
 (Theorem~\ref{th:convpartial}).
We do not give a detailed proof of  this
upper bound but state an auxiliary lemma (Lemma~\ref{th:degraded}) and sketch how this leads to the theorem.

We start by introducing the concept of degradedness for interference
networks with $K_T$ transmitters and $K_R$ receivers. Here, we allow
the number of transmitters to differ from the number of receivers.
Also, in this section we use the concept of multi-antenna interference
networks, that is, we assume that Transmitter~$k$ consists of $t_k$
transmit antennas and Receiver~$j$ consists of $r_j$ receive antennas.
We denote Transmitter~$k$'s time-$t$ channel input by the vector
$\vect{X}_{k}(t)\in \Reals^{t_k}$ and Receiver~$j$'s time-$t$ channel
output by the vector
$\vect{Y}_{j}(t)\in \Reals^{r_j}$. The message sets $\{\set{S}_k\}$
are defined as for the $K$-by-$K$ single-antenna networks. We say that
 an input distribution is allowed if for any time $t$ the vector
 $\vect{X}_k(t)$, $k\in\set{K}$, depends only on Messages $M_j$ for which
                                $j\in \set{S}_k$.% where $t_k$ is the number of
%transmit antennas at Transmitter~$k$ and $r_j$ is the number of
%receive antennas at Receiver~$j$. %By $\vect{Y}_j^n$ we denote the
%sequence of channel output vectors $(\vect{Y}_{j}(1),\ldots, \vect{Y}_{j}(n))$.

\begin{definition} \label{def:degraded}
A $K_T$-transmitters/$K_R$-receivers multi-antenna interference
network is called
%with channel inputs $\vect{X}_k \in \Reals^{t_k}$, where $k\in
%\{1,\ldots,T\}$ and   $t_k$
%denotes the number of antennas at Transmitter~$k$, and with channel 
%outputs $\vect{Y}_j\in \Reals^{r_j}$, where $j\in\{1,\ldots, R\}$ and $r_j$ denotes the number
%of antennas at Receiver~$j$. The interference network is said to be
\emph{degraded} with respect to the permutation $\boldsymbol{\pi}$ on the set of receivers, $\boldsymbol{\pi}:
\{1,\ldots, K_R\} \rightarrow \{1,\ldots, K_R\}$, if any time $t$
\begin{equation}\label{eq:eq_degraded}\vect{Y}_{\boldsymbol{\pi}(1)}(t) \subseteq \vect{Y}_{\boldsymbol{\pi}(2)}(t) \subseteq \ldots \subseteq
\vect{Y}_{\boldsymbol{\pi}(K_R-1)}(t) \subseteq
\vect{Y}_{\boldsymbol{\pi}(K_R)}(t).
\end{equation}  
\end{definition}
Note that the definition does not depend on the side-information
available at the encoders. It is only a property of the channel.

\begin{lemma}\label{th:degraded}
Consider a $K_T$-transmitters/$K_R$-receivers
multi-antenna interference network which is degraded with respect to some permutation
$\boldsymbol{\pi}: \{1,\ldots,K_R\}\rightarrow \{1,\ldots, K_R\}$.
If for all time instants $t$ and for any given allowed input distribution the channel outputs for $j\in \{2,\ldots, K_R\}$ fulfill
\begin{equation}\label{eq:cond_function}
\vect{Y}_{\boldsymbol{\pi}(j)}(t) = f_j\left(
\vect{Y}_{\boldsymbol{\pi}(1)}(t),\ldots,\vect{Y}_{\boldsymbol{\pi}(j-1)}(t),
M_1,\ldots, M_{j-1}\right)%\quad  j\in\{2,\ldots, K_R\}%
\end{equation}
for some set of deterministic functions $\{f_j(\cdot)\}$, then
 the capacity region of the interference network equals the
capacity region of a multi-antenna $K_T$-transmitters/$K_R$-receivers
 interference network where at time $t$ all receivers
observe only the output $\vect{Y}_{\boldsymbol{\pi}(1)}(t)$.
\end{lemma}
The proof of the lemma is omitted. It relies on the fact
that from the channel output sequence %$\mat{Y}^n_{\bfpi(1)}\triangleq
%(
$\vect{Y}_{\bfpi(1)}(1),\ldots,
\vect{Y}_{\bfpi(1)}(n)$ it is possible to reconstruct the sequences
%$\mat{Y}_{\bfpi(2)}^n$
$(\vect{Y}_{\bfpi(2)}(1),\ldots,\vect{Y}_{\bfpi(2)}(n)), \ldots,
(\vect{Y}_{\bfpi(K_R)}(1)$
, $\ldots,$% \mat{Y}_{\bfpi(K_R)}^n$ 
$\vect{Y}_{\bfpi(K_R)}(n))$ 
with 
probability of error tending to 0 for increasing block-lengths $n$
whenever the  rate tuple $(R_1,\ldots, R_{K_R})$ is achievable in
the original network.
%\begin{remark}
%Note that the capacity region of a degraded interference network
%fulfilling Condition~\eqref{eq:cond_function} has the same capacity
%region as multi-access channel with $K_T$ multi-antenna transmitters
%and with one receiver observing at time $t$ $\vect{Y}_{\bfpi(1)}$.
%\end{remark}

Lemma~\ref{th:degraded} is a main tool in the proof of the upper
bound in Theorem~\ref{th:convpartial} below. Before stating the
theorem we want to give a brief outline of how
Lemma~\ref{th:degraded} can be used to prove  an upper bound for %proof.% of Theorem~\ref{th:convpartial}. %This
%should help the reader to better understand the technical
%Condition~\eqref{eq:convpartialcond} in Theorem~\ref{th:convpartial}.
%In a first step we sketch a method on how to obtain  an upper bound on
%the capacity region using Lemma~\ref{th:degraded} for $K$-transmitters/$K$-receivers
interference networks fulfilling a certain technical condition, a
special case of Condition~\eqref{eq:convpartialcond}. After the
statement of the theorem we outline how the proof can be extended to
the more general networks fulfilling Condition~\eqref{eq:convpartialcond}.
%Then, we state the theorem, followed by an outline how this method can
%be adapted to prove the upper in the theorem.
%bound in \eqref{eq:partialconvbound} for networks fulfilling
%Condition~\eqref{eq:convpartialcond}. 

We turn back to consider $K$-transmitters/$K$-receivers interference networks.
A general such network can easily be converted into a degraded
network by choosing an arbitrary permutation $\bfpi$ on the set of
receivers $\set{K}$ and by letting a genie reveal channel outputs
$\vect{Y}_{\bfpi(1)}^n$ through $\vect{Y}_{\bfpi(j-1)}^n$ to
Receiver~$\bfpi(j)$, $j\in\set{K}$. 
Additionally, let a genie reveal
linear combinations $\vect{\tilde{Z}}_{1}^n,\ldots,
\vect{\tilde{Z}}_{K-1}^n$ of the Gaussian noise sequences
$\vect{Z}_1^n,\ldots, \vect{Z}_K^n$ to all receivers. 

Note that these two steps can only increase the sum-rate
 capacity of a network. Therefore, any upper bound on the sum-rate
 capacity of the ``genie-aided'' network is also an upper
 bound on the sum-rate capacity of the original network.

In the sequel, we only consider 
interference networks for which one can choose
a permutation $\bfpi$  and linear combinations
 $\vect{\tilde{Z}}_1^n,\ldots, \vect{\tilde{Z}}_{K-1}^n$ such that for some coefficients $\{\alpha_{j,\ell}\}$ the difference $\vect{Y}_{\bfpi(j)}^n-
\sum_{\ell=1}^{j-1} \alpha_{j,\ell} \vect{Y}_{\bfpi(\ell)}^n$, for $j=2,\ldots,K$, is a function of the Messages
$M_{\bfpi(1)},\ldots, M_{\bfpi(j-1)}$ and the Gaussian sequences
$\tilde{\vect{Z}_1^n},\ldots, \vect{\tilde{Z}}_{K-1}^n$ only. One can
directly  verify that the 
 ``genie-aided'' versions of the networks under consideration fulfill Condition~\eqref{eq:cond_function}
and thus Lemma~\ref{th:degraded} can be applied. We conclude that for the
considered networks any upper bound on the sum-rate capacity (and thus
on the pre-log) of the network where all receivers observe $\vect{Y}_{\bfpi(1)}$ and
$\vect{\tilde{Z}}_1^n,\ldots, \vect{\tilde{Z}}_{K-1}^n$ is also an upper bound
on the sum-rate capacity (and thus on the pre-log) of the original network.
Finally, note that the pre-log of an interference network where all receivers observe
only one
``input-dependent'' output $\vect{Y}_{\bfpi(1)}$ and  Gaussian
noise sequences
$\vect{\tilde{Z}}_1,\ldots, \vect{\tilde{Z}}_{K-1}^n$ which are
correlated with the noise sequence corrupting $\vect{Y}_{\bfpi(1)}^n$ (but
do not determine it)
equals 1.
%Note that we simplified the problem, since the remaining task
%of finding the sum-rate capacity (or an upper bound on it) for a
%network where all receivers observe the same output, is better
%understood. (In fact it corresponds to a  multi-access
%scenario with partially informed transmitters.)

%% Of course, only a very limited number of interference networks have
%% the described structure and thus this method of finding an upper bound
%% on the pre-log is only of use to a small class of networks.
One can extend this method to a larger class of networks, namely those
fulfilling Condition~\eqref{eq:convpartialcond}:
%, from which the
%following theorem follows.

\begin{theorem}\label{th:convpartial}
Consider a $K$-transmitters/$K$-receivers interference network with channel matrix
$\mat{H}$---consisting of rows $\trans{\vect{h}}_1, \ldots, \trans{\vect{h}}_K$--- %$, $j\in \set{K}$,
 and message sets $\{\set{S}_k\}$ where $q+\nu$ distinct rows of
 the channel matrix $\trans{\vect{h}}_{\j_1},\ldots, \trans{\vect{h}}_{\j_q},
 \trans{\vect{h}}_{v_1},\ldots, \trans{\vect{h}}_{v_\nu}$
for any time $t$ fulfill
\begin{equation}\label{eq:convpartialcond}
\left(\trans{\vect{h}_{\j_i}}  - \sum_{\ell=1}^{|\set{V}|} \alpha_{i,\ell}
\trans{\vect{h}}_{v_\ell}\right) \vect{X}(t)
   \indep (M_{\j_i},\ldots, M_{\j_q}), \quad j=1,\ldots,q 
\end{equation}
for some coefficients $\{\alpha_{i,\ell}\}_{\substack{i=1,\ldots,q\\
    \ell =1,\ldots,\nu }}$ and any allowed input distribution. (Here  $\indep$ denotes independence.) % and for any distribution on $\vect{X}(1),\ldots, \vect{X}(n)$
%for which $X_k(1),\ldots, X_k(n)$ depend only on the Messages $M_j$
%for which $j\in\set{S}_k$, $k\in \set{K}$. 
Then, any rate tuple $(R_1,\ldots, R_K)$ can only be achievable if %the rates $R_{j_1},\ldots,
%R_{j_q},R_{v_{1}},\ldots, R_{v_{|\set{V}|}}$ can only be achievable if 
\begin{IEEEeqnarray}{rCl}\label{eq:partialconvbound}
\sum_{i=1}^q R_{\j_i} + \sum_{\ell=1}^{|\set{V}|}
  R_{v_\ell} \leq  \frac{|\set{V}|}{2}\log \left(1+ 
    \|\mat{H}\|^2 P \right)+ \textnormal{c}(\{\alpha_{i,\ell}\})
\IEEEeqnarraynumspace
\end{IEEEeqnarray} 
where $\|\mat{H}\|$ denotes the operator norm of the matrix
$\mat{H}$ and c$(\{\alpha_{i,\ell}\})$ is a constant depending  on the
coefficients $\{\alpha_{i,\ell}\}$.
\end{theorem}

%For theleading to the 
To obtain the 
upper bound in Theorem~\ref{th:convpartial},  adapt
Definition~\ref{def:degraded} to
be applicable also when only a subset of receivers fulfills \eqref{eq:eq_degraded}
for some permutation on this subset, and modify
Lemma~\ref{th:degraded} to apply also for subsets of receivers. %% The steps leading to an upper bound for networks fulfilling
%% \eqref{eq:convpartialcond} are then described as follows:
%The method for obtaining the upper bound then becomes:
Furthermore: Join
Receivers $v_1,\ldots, v_\nu$ into a big common Receiver
$v_{\set{V}}$, thus transforming the $K$-transmitters/$K$-receivers
network into a $K$-transmitters/$(K-\nu+1)$-receivers network; Let a
genie reveal Messages $M_j$, for $j\notin \left(\{v_1,\ldots, v_\nu\}
  \cup\{\j_1,\ldots, \j_q\} \right)$ to Receivers
$v_{\set{V}},\j_1,\ldots,\j_q$; Choose the permutation $\bfpi$:
$\bfpi(1)=v_\set{V}, \bfpi(2)=\j_1,\ldots,\bfpi(q+1)=\j_q$ on the
subset of Receivers  $\{v_{\set{V}},\j_1,\ldots,\j_q\}$;
Let a genie reveal outputs
$\vect{Y}_{v_1}^n,\ldots, \vect{Y}_{v_\nu}^n,
\vect{Y}_{\j_1}^n,\ldots, \vect{Y}_{\j_{i-1}}^n$ and some properly chosen linear
combinations
$\vect{\tilde{Z}}_1^n,\ldots,\vect{\tilde{Z}}_{q}^n$ of  the noise
sequences
$\vect{Z}_{v_1}^n,\ldots, \vect{Z}_{v_\nu}^n, \vect{Z}_{\j_1}^n,\ldots,\vect{Z}_{\j_q}^n$ to
Receiver $\j_i$, $i=1,\ldots, q$; %%  The subset of Receivers
%% $v_{\set{V}},\j_1,\ldots,\j_q$ is degraded with respect to the
%% permutation $\bfpi$. Thus,
Apply the modified version of
Lemma~\ref{th:degraded} to the subset of receivers
$v_{\set{V}},\j_1,\ldots, \j_q$; Finally, derive an upper bound for the
resulting network where Receivers $v_{\set{V}},\j_1,\ldots,\j_q$
observe the same output.

\subsubsection{An Improved Upper Bound}
For an improved upper bound that applies our techniques \emph{simultaneously}
to subsets of the rate, please see a forthcoming longer version of
this paper. 
Some of the results in this paper rely on this improved
upper bound.


\begin{thebibliography}{1}
%\bibitem{telatar99}
%E. Telatar, ``Capacity of multi-antenna Gaussian channels,''
%\emph{European Trans. on Telecomm. ETT}, vol.\ 10, pp.~585--596, 1999;

\bibitem{yoogoldsmith06}
T. Yoo and A. Goldsmith, ``On the Optimality of Multi-Antenna
Broadcast Scheduling Using Zero Forcing Beam Forming,'' 
\emph{IEEE JSAC, Special Issue on 4G Wireless Systems,} Vol. 24,
No. 3, pp. 528--541, March 2006.

 
\bibitem{etkintsewang07}
R. H. Etkin, D. N. C. Tse, and H. Wang, ``Gaussian Interference
Channel Capacity to Within One Bit'', Submitted to \emph{IEEE
  Trans. of Inform. Theory}, 2007.


\bibitem{jafarfakhereddin07}
S. A. Jafar and  M. Fakhereddin, ``Degrees of Freedom on the MIMO Interference Channel,''
To appear in \emph{IEEE Trans. of Inform. Theory}, 2007;

\bibitem{maddahetal06}
M. A. Maddah-Ali, A. S. Motahari, and A. K. Khandani, `` Signaling
over MIMO Multi-base Systems: Combination of Multi-access and
Broadcast Schemes,'' ISIT 2006, Seattle, Washington, July 9-14, 2006.

\bibitem{devroyesharif07}
N. Devroye and M. Sharif, ``The multiplexing gain of MIMO X-channels
with partial transmit side information,'' ISIT 2007, Nice, June 24-29, 2007.

%\bibitem{jafar07}
%S. A. Jafar, ``Degrees of Freedom on the MIMO X Channel'', Preprint,
%available at the author's website.


\bibitem{lapidothshamaiwigger07}
A. Lapidoth, S. Shamai (Shitz), and M. A. Wigger, ``A Linear
Interference Network with Local Side-Information,'' ISIT 2007, Nice, June 24-29, 2007.

\bibitem{weingartenshamaikramer07}
H. Weingarten, S. Shamai (Shitz), and Gerhard Kramer, ``On the
Compound MIMO Broadcast Channel'', Proc.  2007 Workshop on Inform.
Theory and its App., UCSD Campus, La Jolla, CA, Jan.29-Feb.2, 2007.


\bibitem{wyner94}
A. D. Wyner, ``Shannon-Theoretic Approach to a Gaussian Cellular
Multiple-Access Channel'', \emph{IEEE Trans. of Inform. Theory},
Vol. 40, No. 6, November 1994.



%\bibitem{hornjohnson}
%R. A. Horn and C. R. Johnson, ``Matrix Analysis'', Cambridge
%University Press, 1985.

\end{thebibliography}
\end{document}